\documentclass[conference]{IEEEtran}
\usepackage{fancyhdr}

\usepackage{graphicx}
\usepackage{amsmath}
\usepackage{amssymb}
\usepackage{booktabs}
\usepackage{cite}

\usepackage{xspace}
\usepackage{xcolor}
\usepackage{listings}
\usepackage{lstlinebgrd}
\usepackage{siunitx}
\usepackage{tikz}

\usepackage{hyperref}
\usepackage[capitalise,noabbrev]{cleveref}

\usetikzlibrary{arrows.meta, positioning, shapes.geometric}

\lstdefinelanguage{CIVL}[]{C}{
  morekeywords={$input,$assume,$assert},
  alsoletter={\$}%
}

\newcommand{\codefont}{%
  \ttfamily
  \fontsize{6.5pt}{7.5pt}\selectfont
}

\makeatletter
\long\def\lst@makecaption#1#2{%
  \def\@captype{lstlisting}%
  \@makecaption{#1}{#2}%
  \par\vskip\belowcaptionskip\relax}
\makeatother

\definecolor{diffadd}{RGB}{0,120,0}
\definecolor{diffrem}{RGB}{170,0,0}

\newcommand{\flashx}{\mbox{Flash-X}\xspace}
\newcommand{\FLASH}{\mbox{FLASH}\xspace}

\xspaceaddexceptions{’}

\begin{document}

\title{Developing a Numerical Algorithm with CIVL Model Checking in the Loop}

\author{
\IEEEauthorblockN{Youngjun Lee}
\IEEEauthorblockA{\textit{Argonne National Laboratory} \\
Lemont, Illinois, USA \\
leey@anl.gov}
\and
\IEEEauthorblockN{Anshu Dubey}
\IEEEauthorblockA{\textit{RIKEN Center for Computational Science} \\
Kobe, Japan \\
anshu.dubey@riken.jp}
\and
\IEEEauthorblockN{Jan H{\"u}ckelheim}
\IEEEauthorblockA{\textit{Argonne National Laboratory} \\
Lemont, Illinois, USA \\
jhueckelheim@anl.gov}
}

\maketitle
\thispagestyle{fancy}
\lhead{}
\rhead{}
\chead{\scriptsize{\copyright~2026 IEEE. Personal use of this material is permitted.  Permission from IEEE must be obtained for all other uses, in any current or future media, including reprinting/republishing this material for advertising or promotional purposes, creating new collective works, for resale or redistribution to servers or lists, or reuse of any copyrighted component of this work in other works.}}
\lfoot{}
\rfoot{}
\cfoot{}
\renewcommand{\headrulewidth}{0pt}
\renewcommand{\footrulewidth}{0pt}

\begin{abstract}
Verifying a numerical algorithm in a large scientific simulation framework is challenging: the framework is too big to model-check, and the unit tests exercise only sampled inputs. We report a case study in which we developed a new cloud-in-cell (CIC) deposition algorithm for Flash-X, a large-scale multiphysics simulation framework, keeping the CIVL model checker in the development loop. Rather than verify the algorithm within Flash-X’s hefty infrastructure, we extract only the interfaces that the algorithm needs into a small, self-contained C model, which abstracts away implementation details of the Flash-X infrastructure that are unrelated to the new algorithm. The new algorithm is then built and checked within this C model. The CIVL model checker enables verification of the required physical properties of the CIC deposition algorithm using symbolic values for the particle positions. It proves two physical properties---mass conservation and the deposition location---for the continuum of admissible positions in the simulation domain. CIVL also verifies the algorithm’s memory safety and freedom from MPI deadlocks and data race conditions over all rank distributions within specified bounds. Writing the verifying properties first and continuously checking them at each stage of the bottom-up prototyping workflow turned CIVL into a design guardrail that greatly increased confidence in the extended algorithm. During our case study, CIVL surfaced a concurrency defect in the algorithm that our random-seed-based tests failed to exercise. This paper shows our workflow, with the goal of helping readers understand its benefits, cost, and tradeoffs compared with testing.
\end{abstract}


\section{Introduction}
Numerical simulation software is notoriously difficult to get right. A scientific simulation code combines two error-prone concerns at once. The numerical algorithm relies on floating-point operations over a discretized domain, where a subtle bug in calculating an index, a weight, or a boundary case silently degrades the solution rather than crashing. Also, the parallel algorithm distributes the execution over many processes whose interleavings, message orderings, and certain floating-point operations such as summation depend on the process counts, schedules, and system-specific or nondeterministic effects. Conventional test suites address both concerns only by sampling: a test fixes concrete inputs and observes a handful of executions so that it can cover neither the continuum of admissible numerical inputs nor the combinatorial space of parallel schedules and process counts. Consequently, defects that surface only for certain inputs or process distributions can easily slip through. Formal verification, by contrast, promises exhaustive guarantees, but modern simulation codes are usually surrounded by large, layered frameworks, making it hard to verify an entire application stack in place.

CIVL (Concurrency Intermediate Verification Language)~\cite{siegel2015civl} is a verification framework that combines model checking with symbolic execution, in which inputs are taken as symbolic values and the feasibility of each execution path, together with the assertions placed by the developer, is discharged to SMT (Satisfiability Modulo Theories) solvers. CIVL provides front ends for several concurrency models, including MPI, OpenMP, and CUDA, lowering them to a single intermediate language, and it detects deadlocks, data races, illegal memory access, and violations of contracts written by developers. Crucially for numerical code, it can model numerical floating-point computations symbolically as real numbers, enabling reasoning over a continuum of input space rather than a single sampled value. While this model precludes the analysis of floating-point errors, it is a practical approach to reason about the correctness of parallel algorithms, which are often not expected to be bitwise exact, for example due to the non-associativity of floating-point summation. This design choice has allowed CIVL to be applied to production-grade numerical frameworks such as PETSc~\cite{dhavala2025verifying}. 

There is a substantial body of work applying formal methods to message-passing software, mostly concentrated on the parallel-communication layer~\cite{gopalakrishnan2011formal}. Dynamic verifiers such as ISP~\cite{vakkalanka2008isp} replay a program under carefully chosen schedules to expose deadlocks and mismatched communications. MPI-SV~\cite{chen2020mpi} pairs symbolic execution with model checking to reason about MPI programs over symbolic data. Most of this work focuses on verifying a parallel program already established, and the property under test is either communication safety or equivalence to a reference serial implementation.

This paper is a case study in using the model checker as a design-time instrument, not only as a post-hoc verifier. We develop a new particle deposition algorithm intended for \flashx, a large multiphysics simulation framework, keeping CIVL in the development loop from the outset. We build and verify the algorithm in a compact, self-contained model that abstracts the surrounding frameworks of \flashx. We use CIVL not only to certify a finished implementation, but to guide the algorithm’s design by defining correctness properties before the code exists, and re-establishing them at each stage of bottom-up prototyping. We report the detailed workflow that we applied, a concurrency defect exposed by this workflow, and the costs of CIVL model checking in this process.

\section{Background}\label{sec:background}

\subsection{\flashx}\label{sec:background-flashx}

\flashx~\cite{dubey2022flash} is a composable, multiphysics simulation code used in a wide range of scientific domains, enabling users to construct simulations by combining available physics units in \flashx. Most of the physics solvers in \flashx are based on the Eulerian method, which assumes a block-structured Adaptive Mesh Refinement (AMR) grid; however, some physics units are based on the Lagrangian method. The two differ in how they represent matter: Eulerian methods discretize it as field values on a mesh or grid, while Lagrangian methods represent it as moving parcels at continuous positions anywhere in space, such as particles, free of a grid. This coexistence poses a challenge in ensuring physical consistency between the two different pictures of physics modeling. For example, if a simulation is configured to solve the dynamics of free particles on the AMR grid, the physical quantities carried by the particles (e.g., mass, charge) should update the corresponding physical quantities in the grid (e.g., mass density, charge density) to be consistent with the other physics solvers.

The AMR grid in \flashx is organized as a forest of blocks: the domain is decomposed into logically Cartesian blocks, each of which holds a fixed number of interior cells surrounded by a halo of guard cells. The guard cells hold copies of data owned by neighboring blocks, so the stencil-based physics solvers can be performed block-locally, without communicating with blocks that different MPI processors may own. \flashx assumes octree-based AMR, where a block is recursively bisected in every dimension into $2^{d}$ child blocks when a finer resolution is needed. \flashx’s Grid unit handles refinement and derefinement (coarsening) of blocks, along with the data prolongation and restriction between levels, during the simulation runtime; hence, the number of blocks and their topology are undefined before the simulation. \flashx’s physics units apply their numerical algorithm in one block of the AMR grid, considering a block as the finest data-abstraction unit provided by \flashx’s Grid infrastructure.

This infrastructure is the core idea of making \flashx a productive, modular, multiphysics simulation framework, but also makes it difficult to verify the algorithms formally. The Grid unit and its guard-cell exchanges are written in approximately 150k lines of Fortran code with intricate parallel communication, so verifying a new algorithm in place would mean exploring every code path in the entire stack. Crucially, however, checking the physical correctness of an algorithm does not require verifying the internal implementation of \flashx’s Grid unit---it requires only that the algorithm uses the Grid unit’s APIs correctly.

\subsection{Particle-in-Cell Algorithm}\label{sec:background-pic}

In a particle-in-cell (PIC) method, Lagrangian particles move through the Eulerian grid, and the two representations are coupled in each simulation time step: grid quantities are mapped through interpolation, and particle quantities are deposited back onto the grid. A common deposition scheme is the cloud-in-cell (CIC) method~\cite{hockney1988computer}, where each particle is considered a cloud of finite extent, and its quantities are distributed across the surrounding cells with weights proportional to overlap. Therefore, particle deposition is performed on a small, multidimensional stencil, rather than on a single cell. This paper concerns the CIC deposition scheme in \flashx.

\begin{figure}
    \centering
    \includegraphics[width=0.55\linewidth]{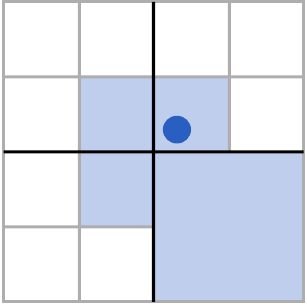}
    \caption{A close look at uneven cloud deposition at a fine-coarse boundary, caused by the reverse-guard-cell-fill-based method}\label{fig:skewed-cloud}
\end{figure}

When a particle sits near a block boundary, its CIC stencil straddles multiple blocks, so its block-local deposition contributes to its guard cell regions. Unlike other stencil-based algorithms in \flashx, where guard cell data is acted upon as temporary copies of neighbor data and then discarded, CIC deposition writes into the guard cells; those updates carry the deposited quantity that belongs to neighboring blocks and must be folded back into their interior cells. In the classical scheme used by \FLASH~\cite{fryxell2000flash}, a predecessor of \flashx, the particle deposits a CIC stencil on its block, then applies a so-called \emph{reverse guard cell fill} process to transfer guard cell data back into interior cells in the neighboring blocks. This works; however, \emph{reverse guard cell fill} poses two issues. One is that the operation is global and communication-heavy, which makes it inefficient. However, it is the second issue that is a greater concern. When a particle is located on a finer cell adjacent to a coarser cell, the CIC mapping gets asymmetric, as demonstrated in \cref{fig:skewed-cloud}. This is because on the fine side the guard cell is only half the size of the coarse cell. During reverse guard cell fill, a prolongation step is applied on the coarse side, and the deposition occurs on an area that is $2^d$ times as big as the original deposition, where $d$ is the dimension of the grid. This changes the shape of the deposited cloud and can introduce spurious forces.

To avoid the spurious forces, we developed an alternative deposition scheme based on \emph{virtual particles}~\cite{dubey2025adding} that replaces the reverse guard cell fill with virtual particle migration. When a particle’s CIC stencil straddles the guard cell regions, we emit one virtual particle per neighboring block the stencil touches, each of which carries the depositing quantity and a position expressed in its destination block’s frame. Virtual particles are then migrated to the MPI ranks that own their destination blocks. After the migration, every particle---including real and virtual particles---can be deposited entirely locally in its owning block, and any deposition in the guard cell region is discarded. Once the deposition is done, the virtual particles are destroyed. The new algorithm also has the virtue of completely eliminating the additional communication step required for reverse guard cell fill. At every timestep, there is a necessary communication step to migrate the particles to their new destination as a result of their physical displacement during evolution. We create and populate virtual particles prior to this migration so that one communication step serves both goals---getting the particles to the right place and ensuring correct deposition at processor boundaries.

\begin{figure}
    \centering
    \includegraphics[width=0.8\linewidth]{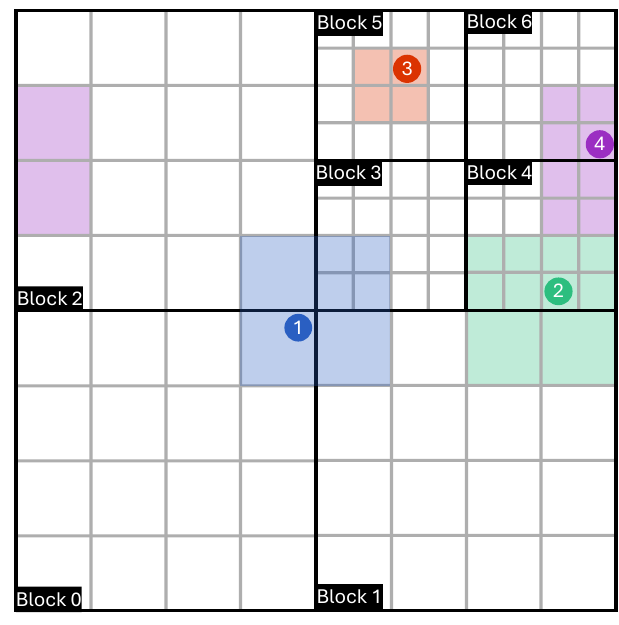}
    \caption{An example AMR grid with periodic boundary conditions for all axes, with four particles. Each particle's CIC stencils are highlighted in each color. All blocks have $4 \times 4$ cells, but have two different refinement levels.}
    \label{fig:amr-particle-stencil}
\end{figure}

In the generation of virtual particles, two families of edge cases must be handled:
\begin{enumerate}
    \item \textbf{Domain boundary condition:} when a virtual particle is generated at a domain boundary, the boundary conditions are resolved at generation time---a periodic edge shifts the deposit position by the domain width, a reflecting edge mirrors it about the boundary, and an outflow edge drops the out-of-domain virtual particle.

    \item \textbf{Fine-coarse interface:} when a virtual particle is generated at an interface between blocks at different refinement levels, the physical size of its CIC stencil must be consistent across the interface---it is resolved at the \emph{coarsest} refinement level that the CIC stencil touches. If a real particle sits in a coarser block, the virtual particle sent to a finer neighbor is spread over $2^{d}$ finer cells (e.g., \cref{fig:amr-particle-stencil}, Particle~1); if the real particle sits in a finer block, its stencil is coarsened to the neighbor’s level (e.g., \cref{fig:amr-particle-stencil}, Particle~2).
\end{enumerate}
In \cref{fig:amr-particle-stencil}, for example, the CIC stencils of four particles are illustrated in an AMR grid with periodic boundary conditions. \textbf{Particle~1} is placed in a block boundary that jumps the refinement level. Three virtual particles are generated and transferred to Blocks 1, 2, and 3. The virtual particle in Block 3 needs to be deposited into four cells instead of one, as the real particle’s and the virtual particle’s refinement levels differ. \textbf{Particle~2} shares a similar circumstance, but the virtual particle’s refinement level is \emph{coarser}. In this case, the real particle’s stencil is extended to match the stencil size from the coarser block. \textbf{Particle~3} lies in a block interior, and its stencil is entirely within one block; thus, no virtual particle is generated---the common case. \textbf{Particle~4}’s stencil spills to the domain’s periodic boundary; so, its virtual particle is generated in Block 2.

Regardless of the AMR geometric complications, a correct deposition must satisfy two physical conditions. Let $\mathcal{D}$ denote the full pipeline under test---virtual particle generation, inter-rank migration, and fine-coarse resolution---and let a particle at position $\mathbf{p}$ in the simulation domain $\Omega$ carry a scalar quantity $q$ (e.g., mass or charge), of which $\mathcal{D}$ deposits $q$'s density, $\rho_{k}$, in cell $k$, whose center is $\mathbf{c}_{k}$ and volume is $\Delta V_{k}$. Then,
\begin{align}
    \sum_{k} \rho_{k} \, \Delta V_{k} &= q, \label{eq:cons} \\
    \sum_{k} \mathbf{c}_{k} \, \rho_{k} \, \Delta V_{k} &= q \, \mathbf{p}. \label{eq:loc}
\end{align}
\Cref{eq:cons} ensures \emph{conservation}: the deposited quantity integrated over the grid equals the quantity carried by the particle, so the operations in $\mathcal{D}$ neither create nor destroy the physical quantity. \cref{eq:loc} checks the \emph{deposition location}: the first moment (e.g., center of mass) of the deposited field equals the particle's position, so the deposited quantity lands in the correct location. The two conditions are independent of each other---e.g., if the quantity is conserved over $\Omega$ but deposited in the wrong place---thus, \cref{eq:loc} is what catches a misplaced stencil, a wrong destination block, or an incorrect refinement resolution, none of which \cref{eq:cons} alone would detect.

Both identities should hold for \emph{every} admissible position $\mathbf{p} \in \Omega$~\footnote {Strictly speaking, \cref{eq:loc} holds in the domain interior, away from the boundaries where a boundary condition intentionally relocates the quantity.}, and under every possible grid configuration that algorithm $\mathcal{D}$ should handle. However, the admissible positions, $\mathbf{p}$, form a continuum, and the possible AMR geometry layered on top of it leads to a large number of corner cases that are difficult to exhaustively explore with a conventional test suite. Additional complications arise because the usual \flashx test mechanism compares against previously validated baselines, but this test practice does not help for algorithmic changes that intentionally change the solution, for example to increase fidelity, as is the case here.
This is what makes the algorithm a natural target for model checking: rather than sample particles' positions, we treat them as symbolic real values and let CIVL prove both conditions by exhaustively exploring all possible branches and message interleavings within bounds.

\section{Method}\label{sec:method}
Our goal is to verify the deposition algorithm \emph{before} it is ported into \flashx, to isolate the algorithm under test from \flashx’s infrastructure. The \flashx framework is too large to model-check, and the algorithm to test does not exist inside it; thus, we reproduce only what the algorithm needs, i.e., APIs required to be called by the algorithm. We apply two levels of abstraction to achieve this: a \emph{physical} abstraction replaces \flashx with a small, self-contained C model of the deposition pipeline, $\mathcal{D}$, and a \emph{logical} abstraction replaces the Grid infrastructure that the model calls with a simplified grid model. A CIVL driver then runs the model with a symbolic particle position and asserts two physical identities of \cref{eq:cons,eq:loc} over the entire space of admissible positions.

Despite \flashx being written in Fortran, we chose C as the language for the abstraction layers, as CIVL's support for C is more mature. Because CIVL has Fortran support~\cite{wenhao2022verifying}, we plan to migrate our approach to Fortran in future work to allow easier re-use across \flashx components.

\subsection{Abstraction Layers}\label{sec:method-abstract}
We develop the algorithm as standalone C code that models the full pipeline of $\mathcal{D}$---particle data structures, virtual particle generations, collective MPI migration, and block-local particle deposit---outside of \flashx. The CIC method implemented in the C model reflects the full logic and algorithm, so it enables us and CIVL to focus on the physical algorithm and its correctness without exploring the surrounding framework. Because the target application, \flashx, runs on distributed nodes, the C model uses the same collective MPI communications.

The only \flashx infrastructure the algorithm depends on is the grid, which it reaches through a small public API: find the block containing a physical position, query the block’s metadata (e.g., refinement level or owning process), and iterate over the block’s data. We implement only that subset of Grid APIs in about 400 lines of C code that reproduces \flashx Grid unit's observable behavior. Although each of the APIs in the simplified grid has a fully functioning implementation, none of them has the production machinery, such as Morton-curve indexing or distribution lookup; they only contain a simplified algorithm that needs to be carried out to check the CIC algorithm. For example, finding a corresponding block’s ID with a given physical coordinate is required for building the CIC algorithm, as the block ID is used for dereferencing the data array to be updated by the algorithm. Leaving the block ID as a symbolic representation cannot supply the concrete value for an array index, so an actual implementation is needed.

This approach allows verification of the algorithm in isolation, while implicitly assuming correctness of the underlying grid data structure. Future work could consider extending this approach to a modular verification of other parts of \flashx, for example by using a spec-stub approach as in~\cite{dhavala2025verifying}.

\subsection{Symbolic Driver}\label{sec:method-civl}
\Cref{lst:civl-driver} shows the CIVL driver, symbolically checking the correctness of the target algorithm. It declares the particle position as a symbolic real input, \texttt{\$input}, constrains it to the domain with \texttt{\$assume}, runs the full pipeline, and checks the conservation and deposition location with \texttt{\$assert}. Since CIVL interprets the \texttt{double} type as real numbers, this single run establishes physical identities \cref{eq:cons,eq:loc} for the entire continuum of physical domain positions, not the finitely many sets of positions that a regular test may handle. The grid topology and the MPI rank counts, by contrast, are concretized and intentionally small. Axiomatizing the AMR grid topology leads to rapid growth in model checking time and requires further optimizations to work within a feasible timeframe, and is left to future work.

The driver deposits a single particle, which is sufficient for the physical identities we are testing. Both \cref{eq:cons,eq:loc} are additive across particles, so proving them for one particle at a symbolic position establishes them for any superposition of particles by linearity. However, the single particle case does not exercise the algorithm's behavior that scales with the particle count. For example, whether a buffer size that is correct for one particle stays correct for many particle cases, or the order in which many particle contributions accumulate in a single cell. However, these aspects can be tested reasonably well with our conventional test suites.

\begin{lstlisting}[
  float=tbp,
  caption={A CIVL driver for the deposition pipeline: a symbolic particle
    position (\texttt{\$input}), constrained by \texttt{\$assume}, checked with
    \texttt{\$assert} for \cref{eq:cons,eq:loc}.},
  label={lst:civl-driver}
]
$input double pos[MDIM]; // symbolic particle position

int main(int argc, char **argv) {
  MPI_Init(&argc, &argv);

  // Initialize a fixed AMR grid
  // Refinement topology is determined by refine_* args
  grid_initStaticAMR(MPI_COMM_WORLD, nb, lo, hi, bc_lo, bc_hi,
      refine_levels, refine_coords, n_refine);

  // Constrain the symbolic position to the domain,*@\;$\texttt{p} \in \Omega$@*.
  double p[MDIM];
  for (int a = 0; a < NDIM; ++a) {
    $assume(pos[a] >= lo[a] && pos[a] <= hi[a]);
    p[a] = pos[a];
  }

  // Initialize one unit-mass particle
  particle_t r = make_real(p, /*mass=*/1.0);
  // Run the full pipeline under test,*@\;$\mathcal{D}$@*
  int n_reals = rank_owns(p) ? 1 : 0;
  particles_deposit(&r, n_reals);

  // *@\Cref{eq:cons}@*: total deposited mass == input mass
  $assert(deposited_mass() == 1.0);

  // *@\Cref{eq:loc}@*: first moment == p, interior only
  double mom[MDIM];
  deposited_moment(mom);
  for (int a = 0; a < NDIM; ++a)
    if (interior(p, a))
      $assert(mom[a] == p[a]);

  grid_finalize();
  MPI_Finalize();
}
\end{lstlisting}

\subsection{Model Checking in the Loop}\label{sec:method-devloop}
The abstraction layers and the symbolic model checking driver with the correctness properties described in \cref{sec:method-abstract,sec:method-civl} are used not only as a post-implementation verifier, but also as a tool in the development process. We first write the correctness properties that the target algorithm must adhere to, so that ``correct’’ is defined before the algorithm code exists, similar to test-driven development. Conventional test suites can also coexist alongside it. Still, the CIVL driver provides two additional correctness checks that test suites cannot: CIVL \emph{proves} the correctness properties over the continuous space, and it checks the model defects such as out-of-bounds access, invalid pointers, and MPI deadlocks and race conditions. The algorithm is then developed to satisfy them, with the properties as \emph{formal contracts} that CIVL checks at every iteration of development. It proves whether the current version is correct on the predefined properties, or returns a counterexample that fails.

We applied this process to build and design the CIC algorithm for \flashx. We first established the CIVL driver in \cref{lst:civl-driver}, then created the simplified grid as an abstraction layer. To develop the complete algorithm, we proceeded in four stages of increasing complexity: a uniform, single-level one-dimensional grid on one rank; then cross-rank in the same grid; then two-dimensions; and finally multi-level AMR. Because all stages share the same correctness properties encoded in the CIVL driver, we re-apply CIVL with the same specification at every stage, extending the algorithm incrementally from a foundation that has already been proved in the previous stage. This workflow helped isolate the counterexample found by CIVL as an implication of the added code in the current stage.

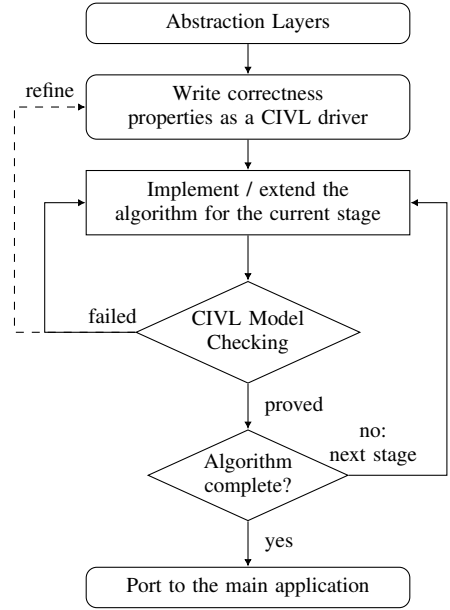
\begin{figure}[t]
\centering
\begin{tikzpicture}[
  font=\footnotesize,
  >={Latex[width=1mm,length=1mm]},
  box/.style ={draw, rounded corners, align=center, inner sep=4pt, text width=40mm},
  proc/.style={draw, align=center, inner sep=4pt, text width=40mm},
  dec/.style ={draw, diamond, aspect=2.2, align=center, inner sep=1pt},
]
\node[box]                       (abst)   {Abstraction Layers};
\node[box,  below=4mm of abst]   (spec)   {Write correctness properties as a CIVL driver};
\node[proc, below=4mm of spec]   (impl)   {Implement / extend the algorithm for the current stage};
\node[dec,  below=6mm of impl]   (verify) {CIVL Model\\Checking};
\node[dec,  below=6mm of verify] (done)   {Algorithm\\complete?};
\node[box,  below=6mm of done]   (port)   {Port to the main application};

\draw[->] (abst)   -- (spec);
\draw[->] (spec)   -- (impl);
\draw[->] (impl)   -- (verify);
\draw[->] (verify) -- node[right=1mm]{proved} (done);
\draw[->] (done)   -- node[right=1mm]{yes} (port);

\draw[->, dashed] (verify.west) -- ++(-16mm,0) |- node[above,near end]{refine} (spec.west);

\draw[->] (verify.west) -- node[above,near start]{failed} ++(-12mm,0) |- (impl.west);

\draw[->] (done.east) -- node[above,near start,align=center]{no:\\next stage} ++(13mm,0) |- (impl.east);
\end{tikzpicture}%
\caption{CIVL-in-the-loop development: each development stage is proved through CIVL.}\label{fig:in-the-loop}
\end{figure}

\Cref{fig:in-the-loop} represents the flowchart described in this section: establish the correctness properties and build the abstraction layers, then develop the algorithm with CIVL in the loop in each stage. CIVL’s check failure usually locates the incorrectness of the algorithm in the current stage, but rarely indicates that the correctness property should be updated/refined. In our case, for example, the deposition location identity, \cref{eq:loc}, does not hold near the domain boundary under non-outflow conditions, as the deposited quantity is intentionally relocated. We found this during the development process and updated our CIVL driver to constrain the deposition location check only when the particle is placed away from the domain boundary. 

The value of the loop is the confidence at each stage of the development. It proves correctness properties and searches for model defects over the whole input space, helping developers design and extend the algorithm stage by stage.

\section{Results and Discussion}\label{sec:results}
Following the workflow of \cref{sec:method-devloop}, we designed and developed the CIC particle deposition pipeline on the simplified grid. CIVL proves the conservation, \cref{eq:cons}, for all particle positions in the grid domain, $\Omega$, and the deposition location, \cref{eq:loc}, holds for all positions in the interior of the grid. In the same runs, CIVL establishes memory safety (e.g., out-of-bounds or invalid pointer dereference) and freedom from MPI deadlocks and data races, over all particle placements at each rank count. We developed the algorithm pipeline from the simplest case, a 1D uniform grid with one MPI rank, progressively extended to 2D AMR with multiple ranks, and CIVL verified the correctness at each stage. We verified the pipeline on two AMR configurations: two-level and three-level, both in 1D and 2D with multiple MPI ranks. Runtime and scaling for these CIVL model checks are reported in \cref{sec:costs}.

CIVL model checking also exposed a defect in the pipeline that our conventional tests had not triggered. A rank that owns no real particles can still receive virtual particles if neighboring ranks have real particles near the block boundaries. The receive buffer was sized based on each rank’s \emph{local} real particle counts with a \texttt{max\_total < 1} clamp, so the receive buffer would have only one slot if there is no real particle in the rank. One slot is enough to receive one virtual particle; however, receiving two or more virtual particles is a reachable state, as the neighboring ranks can have many real particles that spill virtual particles. This is also a reachable state in our CIVL driver, even though it has only one particle in the simulation domain. If a particle generates multiple virtuals, then each of them is transferred to the destination blocks, and they can be co-located in the same rank. A rank that owns multiple blocks may require more than one particle slot for the receive buffer, even though it has no real particle. Our test suites did not capture this corner case, as they initialize about a hundred random particles with a random seed, so every rank owned at least one real particle and the empty-rank case was never reached. The CIVL driver, by contrast, exercises one particle over all positions and reaches this corner case, returning a concrete counterexample. The defect would corrupt memory silently instead of crashing. The subsequent communication step appends each received particle to the local pool without a bounds check, so an under-allocated pool is an out-of-bounds heap write. Consequently, the resulting symptom would be a silent wrong deposition rather than an abort. CIVL reports it deterministically as an out-of-bounds violation, without needing the fault to occur at runtime. We sized the buffer from the global particle count to fix this defect, as represented in \cref{lst:poolfix}.

\begin{lstlisting}[
  float=tb,
  language=C,
  keepspaces=true,
  linebackgroundcolor={%
    \ifnum\value{lstnumber}=2\color{diffrem!15}\fi
    \ifnum\value{lstnumber}=3\color{diffadd!15}\fi
    \ifnum\value{lstnumber}=4\color{diffadd!15}\fi
    \ifnum\value{lstnumber}=5\color{diffadd!15}\fi},
  caption={The fix CIVL prompted: size the receive buffer from the \emph{global}
    particle count, so a rank with no local reals is not under-allocated
    ($V$ = max virtuals per real).
  },
  label={lst:poolfix}
]
 int particles_maxPoolSize(int n_reals) {
-  int max_total = n_reals * grid_getNumProcs() * (1 + V);
+  int global = n_reals;
+  MPI_Allreduce(&n_reals, &global, 1, MPI_INT, MPI_SUM, comm);
+  int max_total = global * (1 + V);
   if (max_total < 1) max_total = 1;
   return max_total;
 }
\end{lstlisting}

In summary, the CIVL-in-the-loop procedure demonstrated its benefits in the practical example: developing the CIC deposition algorithm for \flashx. It acted as a guardrail for a bottom-up prototyping strategy and also surfaced an algorithmic defect. These guarantees, however, hold only for the bounded model we check; the grid configuration that the CIVL verifies is fixed in the driver (e.g., \texttt{grid\_initStaticAMR} call in \cref{lst:civl-driver}), so it checks only one AMR topology at a time. In a real case in \flashx simulation, the AMR grid progressively refines and derefines blocks during runtime, so the algorithm should be tested on a large number of possible AMR grid configurations. We used a finite number of grid configurations for our CIVL driver, and its runtime and scalability will be discussed in \cref{sec:costs}. 

\section{Costs and Limitations}\label{sec:costs}
\Cref{tab:civl-stats-a,tab:civl-stats-b} report the cost of the CIVL runs we used, for the two- and three-level AMR grids, respectively. The \emph{Config} column describes the grid dimensionality and the number of MPI ranks. For instance, 2D $\times$ 2 denotes a 2D grid, and its blocks are distributed over two MPI ranks. \emph{Time} and \emph{Mem} are wall-clock time and peak memory used during CIVL model checking. \emph{States} is the number of distinct program states CIVL explored: the size of the reachable state space it searched; \emph{States saved} is the number of states it stored during the depth-first search for backtracking and revisit detection. \emph{Transitions} represents the number of state-to-state execution steps, and \emph{Prover calls} is the number of queries dispatched to the external SMT solver.

\begin{table}[t]
\centering
\scriptsize
\setlength{\tabcolsep}{2pt}
\caption{CIVL stats on two-level AMR grid}\label{tab:civl-stats-a}
\begin{tabular}{lrrrrrr}
\toprule
Config & {Time (s)} & {States} & {States saved} & {Transitions}
       & {Prover calls} & {Mem (GB)} \\
\midrule
1D $\times$ 1 & \num{10.34}   & \num{16008}   & \num{25276}   & \num{21627}   & \num{547}   & \num{1.13} \\
1D $\times$ 2 & \num{12.72}   & \num{27411}   & \num{43264}   & \num{41644}   & \num{526}   & \num{3.40} \\
1D $\times$ 3 & \num{17.85}   & \num{39063}   & \num{61621}   & \num{61654}   & \num{540}   & \num{3.31} \\
2D $\times$ 1 & \num{1063.35} & \num{915753}  & \num{1481838} & \num{1108136} & \num{34434} & \num{4.50} \\
2D $\times$ 2 & \num{1713.58} & \num{1176850} & \num{1891732} & \num{1540286} & \num{33213} & \num{5.82} \\
\bottomrule
\end{tabular}
\end{table}

\begin{table}[t]
\centering
\scriptsize
\setlength{\tabcolsep}{2pt}
\caption{CIVL stats on three-level AMR grid}\label{tab:civl-stats-b}
\begin{tabular}{lrrrrrr}
\toprule
Config & {Time (s)} & {States} & {States saved} & {Transitions}
       & {Prover calls} & {Mem (GB)} \\
\midrule
1D $\times$ 1 & \num{11.53}   & \num{25073}   & \num{39852}   & \num{33238}   & \num{756}   & \num{1.12} \\
1D $\times$ 2 & \num{16.41}   & \num{40631}   & \num{64146}   & \num{59940}   & \num{741}   & \num{3.31} \\
1D $\times$ 3 & \num{23.28}   & \num{57320}   & \num{90530}   & \num{88347}   & \num{724}   & \num{3.24} \\
2D $\times$ 1 & \num{1909.89} & \num{1771007} & \num{2871930} & \num{2121203} & \num{53978} & \num{3.54} \\
2D $\times$ 2 & \num{2978.64} & \num{2190454} & \num{3519344} & \num{2786977} & \num{52089} & \num{6.48} \\
\bottomrule
\end{tabular}
\end{table}

The tables show a steep cost of grid dimensionality. In one dimension, every configuration completes in about twenty seconds with several hundred prover calls, which is cheap enough to run within development iterations. However, it is far more expensive in two-dimensional cases. The states, transitions, and prover calls grow by roughly $50$--$60\times$ in the two-level AMR grid, as the symbolic particle position explored a larger stencil on more cells/blocks. The runtime grows faster, roughly $100\times$, which indicates the individual solver queries also become harder. We observed that two-dimensional configurations require a slightly longer timeout for the SMT solver in CIVL. Refinement depth compounds the cost of CIVL model checking. For 2D configurations, the three-level AMR grid nearly doubles the runtime and states compared to the two-level grid, whereas it requires about a $40$-- $50\%$ states increase and about $10$--$30\%$ more runtime in 1D configurations. On the other hand, we found that extending the number of MPI ranks is comparatively moderate, under $2\times$ across the rank counts we tested.

The runtime cost of 1D and 2D runs decides where CIVL sits in the development loop. The 1D runs are fast enough to wait on after an edit, similar to compilation and unit test runs, and we used them that way throughout development. The 2D runs, however, are too expensive to run after every small code edit, so we occasionally launched them on major code changes. Consequently, most iteration in the workflow of \cref{sec:method-devloop} happens in 1D, with the 2D configurations serving as a checkpoint at the end of each stage.

As shown in the CIVL runtime tables, even a single-rank 2D AMR configuration is already expensive, and this is the main reason to fix the grid topology in the driver rather than expressing the grid abstraction unit in symbolic representations. Expressing the AMR grid topology with symbolic values or verifying the algorithm for three spatial dimensions is therefore left for future work. 

The guarantees we obtain are relative to the fidelity of the abstractions we used, which is an important limitation of this approach. We verified the algorithm on a self-contained abstracted model, rather than on the target application, \flashx. Therefore, transferring the resulting algorithm to \flashx rests on two assumptions. First, the simplified grid must reliably reproduce the observable behavior of the \flashx Grid APIs that the algorithm calls. We limit this exposure by modeling only the small subset of well-defined public APIs in simplified algorithms/implementations, rather than reimplementing the full functionality the \flashx Grid unit provides. Second, the algorithm is checked in C, but the target application is written in Fortran. Reconstructing the properties in the ported code is required to close the gap, and it is on our roadmap. Even before that port, however, verifying the model pays off: it pins the design of the algorithm and verifies that it is correct on an AMR grid. Thus, porting the abstraction layers and algorithm to Fortran begins from a design that has already proved correct, eventually simplifying the algorithm integration to \flashx.

It is worth noting that establishing and building the abstraction layers is a small one-time cost. The simplified grid model that we used as an abstraction layer is about 400 lines of C code, and the CIVL driver adds about 150 lines of code. More importantly, most of the effort is reusable. The grid abstraction reproduces most of the \flashx Grid API that any \flashx physics unit calls, so it is a reusable component and testbed for a new \flashx physics solver, bringing CIVL into the design and development workflow. Developing another physics solver in \flashx would reuse the same grid model and the same workflow, rewriting only the algorithm’s pipeline and its correctness properties. Therefore, the cost of developing the abstraction layers is amortized not only across the stages of one algorithm but across future algorithms as well. 

Keeping the grid abstraction aligned with \flashx, however, is a continuing obligation rather than a one-time cost. Nothing automatically checks and verifies the observable behaviors of the simplified grid model, since the model and \flashx are separate codebases. We consider this an acceptable trade-off to make model checking feasible, since the maintenance cost follows the handful of Grid APIs needed for physics algorithms under test, not 150k lines of code from the \flashx Grid unit implementation.

\section{Conclusion and Future Work}\label{sec:conclusions}
We presented a case study on developing a new numerical algorithm (the CIC deposition scheme based on the virtual particle concept) for \flashx, keeping the CIVL model checker in the loop. Instead of verifying the algorithm inside \flashx’s large infrastructure, which is impractical to model-check, we introduced abstraction layers that reproduce a small subset of the \flashx framework with minimal implementation. On top of the abstraction layers, the target algorithm is implemented as a small, self-contained C program, enabling the CIVL model checker to exercise the required physical properties with symbolic values. Treating the particle position as a symbolic real value, CIVL proved mass conservation and the first moment of the deposited field over the continuum of particle positions in the simulation domain. Additionally, it checked memory safety and absence of MPI deadlocks and race conditions across the multiple-rank distribution we tested. 

The core of the methodology is writing the correctness properties in CIVL before the algorithm and continuously checking them at every stage of a bottom-up prototyping workflow. This turned CIVL from a post-hoc verifier into a design guardrail. At each stage of development, the algorithm was extended from the foundation already proven through CIVL, and CIVL surfaced the concurrency defect that our conventional test suites failed to exercise. The computational cost for the CIVL model checking is modest in one-dimensional cases, but considerably more expensive in two-dimensional cases. The abstraction layers developed during the workflow are reusable components for future \flashx development, so the model-checking-in-the-loop procedure can be applied to other new physics solver development.

Yet, several directions remain to strengthen the workflow. First, the CIVL driver fixes the AMR grid topology and checks only one AMR configuration at a time. Expressing the topology itself with symbolic values would enable CIVL to verify the target algorithm over all admissible grid configurations; however, as \cref{sec:costs} shows, this multiplies an already non-trivial state space search, which makes it impractical. Additional abstractions and state-reduction techniques are required. Second, we plan to port the abstraction layers and the current algorithm implementation to Fortran, the main language of \flashx. We used C to leverage the most mature support from CIVL, but as reported in~\cite{wenhao2022verifying}, CIVL already has Fortran support. As we port our C abstraction layers to Fortran, we expect to improve and expand CIVL’s Fortran capability, which we hope will help the community in verifying Fortran code. Additionally, the Fortran abstraction layer will be more proximate to \flashx, and we will eventually port the algorithm to \flashx.

\section*{Acknowledgment}
This material is based upon work supported by the U.S.\ National
Science Foundation under Award Number CCF-2446130, and by the U.S.\
Department of Energy, Office of Science, Advanced Scientific Computing
Research (ASCR) Program, under Award Number DE-SC0025953.

{\small This report was prepared as an account of work sponsored by an agency
of the United States Government. Neither the United States Government
nor any agency thereof, nor any of their employees, makes any
warranty, express or implied, or assumes any legal liability or 
responsibility for the accuracy, completeness, or usefulness of any
information, apparatus, product, or process disclosed, or represents
that its use would not infringe privately owned rights. Reference
herein to any specific commercial product, process, or service by
trade name, trademark, manufacturer, or otherwise does not necessarily
constitute or imply its endorsement, recommendation, or favoring by
the United States Government or any agency thereof. The views and
opinions of authors expressed herein do not necessarily state or
reflect those of the United States Government or any agency thereof.}



\bibliographystyle{IEEEtran}
\bibliography{main}

\end{document}